\documentclass[reprint,superscriptaddress,showpacs,preprintnumbers,amsmath,amssymb,floatfix,aps,prl]{revtex4-1}

\usepackage{graphicx}
\usepackage{dcolumn}
\usepackage{bm}

\begin{document}
\title{
Quantum spin fluctuations in the spin liquid state of Tb$_2$Ti$_2$O$_7$
}

\author{Hiroshi Takatsu}
\affiliation{Department of Physics, Tokyo Metropolitan University, Hachioji-shi, Tokyo 192-0397, Japan}

\author{Hiroaki Kadowaki}
\affiliation{Department of Physics, Tokyo Metropolitan University, Hachioji-shi, Tokyo 192-0397, Japan}

\author{Taku J. Sato}
\affiliation{NSL, Institute for Solid State Physics, University of Tokyo, Tokai, Ibaraki 319-1106, Japan}

\author{Jeffrey W. Lynn}
\affiliation{NCNR, National Institute of Standards and Technology, Gaithersburg, MD 20899-6102, U.S.A.}

\author{Yoshikazu Tabata}
\affiliation{Department of Materials Science and Engineering, Kyoto University, Kyoto 606-8501, Japan}

\author{Teruo Yamazaki}
\affiliation{Institute for Solid State Physics, University of Tokyo, Kashiwa 277-8581, Japan}

\author{Kazuyuki Matsuhira}
\affiliation{Department of Electronics, Faculty of Engineering, Kyushu Institute of Technology, Kitakyushu 804-8550, Japan}

\date{\today}

\begin{abstract}
Neutron scattering experiments on a polycrystalline sample of 
the frustrated pyrochlore magnet Tb$_2$Ti$_2$O$_7$, which does not show any magnetic order down to 50~mK, 
have revealed that it shows condensation behavior below 0.4~K from a 
thermally fluctuating paramagnetic state to a spin-liquid 
ground-state with quantum spin fluctuations. 
Energy spectra change from quasielastic scattering to 
a continuum with a double-peak structure at energies of 0 
and 0.8~K in the spin-liquid state. 
Specific heat shows an anomaly at the crossover temperature. 
\end{abstract}

\maketitle

\noindent \textbf{1. Introduction}

Magnetic systems with geometric frustration, 
a prototype of which is antiferromagnetically coupled 
Ising spins on a triangle, have been intensively studied 
experimentally and theoretically for decades \cite{Gardner10,Diep04}. 
Spin systems on networks of triangles or tetrahedra, 
such as triangular \cite{Wannier50}, kagom\'{e} \cite{Mekata03}, 
spinel \cite{Anderson56}, and pyrochlore \cite{Gardner10} lattices, 
play major roles in these studies. 
Their rich variety of phenomena includes 
zero temperature entropy in Ising antiferromagnets \cite{Wannier50,Anderson56} 
and in ferromagnetically coupled spin-ice \cite{Harris97}, 
multiferroics induced by non-collinear magnetic structures \cite{Tokura2010}, 
heavy fermion behavior \cite{Shiga93}, 
and unconventional anomalous Hall effect \cite{Taguchi01,Takatsu10}. 

Subjects that have fascinated many investigators in recent 
years are classical and quantum spin-liquid states \cite{Anderson73,SLP,Lee08,Balents10}, 
where conventional long-range order (LRO) is suppressed 
to very low temperatures. 
Quantum spin-liquids in particular have been 
challenging both theoretically and experimentally 
since the proposal of the resonating valence-bond state \cite{Anderson73}. 
A prototype spin-liquid, or cooperative paramagnet, is 
the Heisenberg antiferromagnetic model on the pyrochlore 
lattice coupled by nearest-neighbor exchange interaction, 
which is shown to remain disordered at all temperatures \cite{SLP}. 
The spin ice materials, R$_2$T$_2$O$_7$ (R = Dy, Ho; T = Ti, Sn), are
the best-understood classical examples \cite{Bramwell01}, 
while other experimental candidates found recently are awaiting further studies \cite{Lee08,Balents10}.

Among magnetic pyrochlore oxides \cite{Gardner10}, 
Tb$_2$Ti$_2$O$_7$ has attracted much attention 
because it does not show any magnetic LRO down to 50 mK 
and remains dynamic with short range correlations \cite{Gardner99,Gardner03}. 
Theoretical considerations of the crystal-field (CF) states 
of Tb$^{3+}$ and exchange and dipolar interactions of 
the system \cite{Gingras00,Enjalran04,Kao03} 
showed that it should undergo a transition into 
a magnetic LRO state at about $T \sim 1.8$ K within a random 
phase approximation \cite{Kao03}.
The dynamical ground state is a candidate for a quantum 
spin-liquid, but its puzzling origin has been in debate \cite{Gardner10}. 

Recently, an interesting scenario to explain this spin-liquid 
state was theoretically proposed~\cite{Molavian07}; 
Tb$_2$Ti$_2$O$_7$ is a quantum-mechanical version of 
the classical spin-ice \cite{Harris97}, 
where additional spin-flip terms to the otherwise classical 
Ising-spin Hamiltonian 
lift the macroscopic degeneracy of the classical 
``2-in, 2-out'' ground states \cite{Molavian07}. 
More recently a single-site mechanism 
was proposed \cite{Bonville11} to account for the absence of LRO, 
in which the CF ground doublet becomes 
two singlets owing to a conjectured tetragonal 
distortion \cite{Rule08,Chapuis10}, 
although this interpretation is not without 
difficulties \cite{Gaulin11}. 

In addition to the theoretical puzzle about the ground state, 
some experimental results of Tb$_2$Ti$_2$O$_7$ contradict 
each other \cite{Gardner99,Gardner03,Yasui02,Hamaguchi04}. 
It was concluded that the majority of 
the spins is dynamic down to 50 mK in \cite{Gardner03}. 
On the other hand, in \cite{Yasui02,Hamaguchi04} 
it was reported that about 50\% of the spins are static at 0.4 K 
and that there is an unknown phase transition at 0.37 K. 
This discrepancy is probably caused by a certain 
uncontrollable parameter of crystalline samples. 
In fact, a recent study of specific heat showed a sample 
dependence for single crystals \cite{Chapuis10,Chapuis09}. 
However by restricting oneself to experimental data 
of polycrystalline samples, results are more consistent. 
Muon spin relaxation ($\mu$SR) \cite{Gardner99}, 
neutron spin echo (NSE) \cite{Gardner03}, 
and susceptibility \cite{Luo01} experiments showed 
no conventional phase transition and LRO. 
Only a small amount ($\sim 10$ \%) of spins become quasi-static 
at 0.1-0.3 K \cite{Gardner03,Luo01}. 
High-resolution neutron powder-diffraction 
experiments~\cite{Han04} showed that the crystal structure 
of Tb$_2$Ti$_2$O$_7$ is consistent with the pyrochlore 
structure without disorder, Tb/Ti site interchange, or 
oxygen deficiency. 

In this work, we hypothesize that polycrystalline samples 
show the genuine characteristics of the spin-liquid state of 
Tb$_2$Ti$_2$O$_7$, 
and reinvestigate the low-temperature spin fluctuations 
of a polycrystalline sample by inelastic neutron scattering. 
From the NSE experiment \cite{Gardner03}, 
we expect that important spin fluctuations of the 
spin-liquid state should appear in the energy range 
$E > 0.05$ meV, i.e., NSE time $<$ 0.01 ns (figure 3 of 
\cite{Gardner03}), where no experimental data have been reported. 
The other aim of this work is to observe a certain temperature 
dependence of energy spectra around $\sim 0.5$ K, anticipated from 
the quantum spin-ice theory \cite{Molavian07}. 
In fact, the NSE \cite{Gardner03} and our unpublished 
neutron scattering experiment \cite{Kadowaki02} 
did suggest such a $T$ dependence. 
We have found that around 0.4~K a high-temperature 
quasielastic spectrum becomes a continuum with a double-peak structure 
at energies of 0 and 0.8~K, indicating that a crossover from 
the paramagnetic to spin-liquid state occurs. 
Specific heat shows an anomaly at the crossover temperature.

\noindent \textbf{2. Experiment}

Polycrystalline samples of Tb$_2$Ti$_2$O$_7$ were 
prepared by the standard solid-state reaction 
at 1350 $^\circ$C from Tb$_4$O$_7$ and TiO$_2$ \cite{Gardner99}. 
Most of the neutron-scattering measurements were
performed on the triple-axis spectrometer 
NG5 at the NIST Center for Neutron Research. 
A sample with a weight of 7 g was mounted in 
a dilution refrigerator. 
The spectrometer was operated using a final neutron 
energy of $E_{\text{f}}=2.5$ meV, providing an 
energy resolution of 0.06 meV (full width at half maximum, 
FWHM) at the elastic position.
Higher-order neutrons were removed by cooled Be and 
BeO filters. 
A few preliminary measurements were performed on the triple-axis 
spectrometer HER at the Japan Atomic Energy Agency \cite{Kadowaki02}. 
Specific heat was measured by the heat-relaxation method 
on a physical-property measurement-system 
equipped with a $^3$He refrigerator. 

\begin{figure}
\begin{center}
 \includegraphics[width=0.45\textwidth]{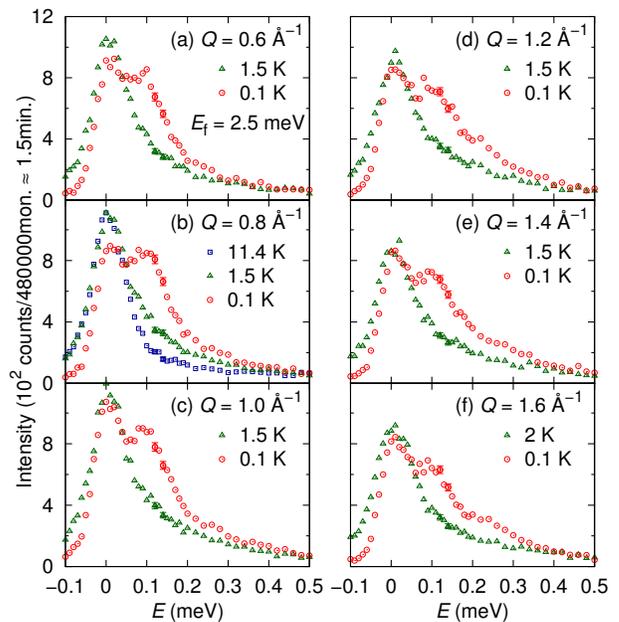}
\caption{
Constant-$Q$ scans taken at 
$Q=0.6$, 0.8, 1.0, 1.2, 1.4, 1.6~\AA$^{-1}$ 
in the energy range $-0.1 \leq E \leq 0.5$ meV 
measured at temperatures above and below 0.4~K.
}
\label{fig.1}
\end{center}
\end{figure}
\begin{figure}
\begin{center}
 \includegraphics[width=0.45\textwidth]{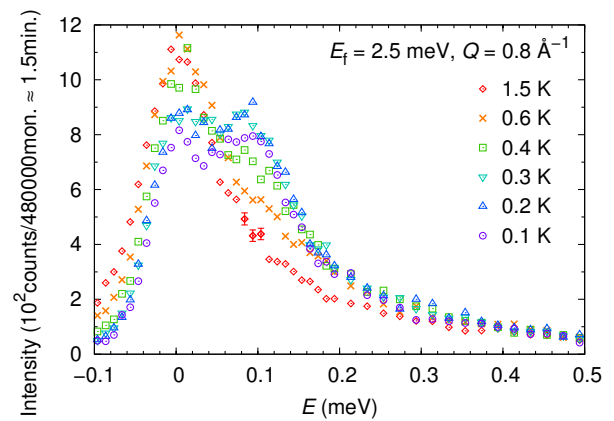}
\caption{
Constant-$Q$ scans at $Q=0.8$~\AA$^{-1}$ in 
an energy range $-0.1 \leq E \leq 0.5$ meV 
measured at several temperatures down to 0.1~K.
}
\label{fig.2}
\end{center}
\end{figure}

\noindent \textbf{3. Results and discussion}

To investigate the ground state, we performed inelastic 
scattering experiments in a low energy range 
$-0.1 \leq E \leq 0.5$~meV at low temperatures. 
In figures~\ref{fig.1} and \ref{fig.2} we show constant-$Q$ scans 
measured at several wave vectors in a range 
$0.6 \leq Q \leq 1.6$~\AA$^{-1}$. 
Since the first excited CF level is about 18~K 
or 1.7~meV \cite{Gardner99,Mirebeau07}, 
the $E$ spectra of these figures are transitions 
among the ground doublet of the CF states of Tb$^{3+}$, being in the trigonal symmetry \cite{Gingras00}. 
The typical energy scale of these $E$ spectra is the order of 
0.1 meV $\sim$ 1 K. 
This value agrees well with the estimate of the effective 
Tb-Tb spin exchange-interaction 
based on $\mu$SR \cite{Gardner99,Gingras00}. 
Thus the observed $E$ spectra represent spin fluctuations of 
up and down states of the Tb$^{3+}$ Ising-like spins \cite{Gingras00}, 
interacting via the inter-site couplings and virtual 
excitations to the excited doublet \cite{Kao03,Enjalran04,Molavian07}. 

One can see from figure \ref{fig.1} that the $E$ spectra change 
remarkably between $T=1.5$ and 0.1~K from a quasi-elastic scattering 
centered at $E=0$ to an inelastic scattering with an additional 
peaked structure at $E \simeq 0.1$~meV. 
Temperature dependence shown in figure \ref{fig.2} indicates that 
the spectral change occurs around $T=0.4$~K. 
We think that this crossover has an important 
implication that the paramagnetic states above and below 0.4~K 
have very different characteristics. 
The $E$ spectra (figure \ref{fig.2}) become a low-$T$ limit 
only below 0.3~K, and we may conclude that Tb$_2$Ti$_2$O$_7$ condenses 
into the spin-liquid ground-state below this temperature. 
It should be noted that in the same crossover $T$ range 
a significant change was observed in the NSE spectra (figure 3 of 
\cite{Gardner03}). 
In contrast to this inelastic scattering, 
the energy-integrated diffraction varies only slightly below 
1.5 down to 0.05~K (figure 2 of \cite{Gardner03}), i.e., 
spin correlations over a single tetrahedron 
are retained at low temperatures \cite{Gardner99}. 

The $E$ spectra in figure \ref{fig.1} above 0.2 meV at 0.1 K exhibit 
additional $Q$-dependent structures. 
This $Q$-dependence seems to impose a restriction on the origin of the spin-liquid 
ground-state; it is brought about by a many-body effect \cite{Gingras00,Kao03,Enjalran04}. 
An interesting proposal along this line is the 
quantum spin-ice state, where a singlet ground state is 
formed predominantly from the ``2-in, 2-out'' classical 
spin-ice states within a single tetrahedron \cite{Molavian07}. 
The estimate of an energy band spanned by exited states 
is the order 0.5 K \cite{Molavian07}. 
This value approximately agrees with the fit parameter $\Delta$ of the inelastic 
Lorentzian function at $T \leq 0.3$~K, which is discussed later.

Quite recently another possibility of the spin-liquid state considering 
both single- and inter-site effects has been pointed out \cite{Bonville11}.
The authors hypothesize that the ground doublet in the cubic pyrochlore 
structure splits into two singlets under a 
small tetragonal distortion, 
conjectured to exist as high as 1.6~K \cite{Bonville11}. 
The energy splitting of 0.19 meV between the two singlets, 
which is claimed \cite{Bonville11} to be observed at 1.6~K 
in a neutron inelastic spectrum in \cite{Mirebeau07}, 
is not reproducible in the present data of figures \ref{fig.1} 
and \ref{fig.2} at 1.5~K. 
The energy resolution 0.16 meV (FWHM) in \cite{Mirebeau07} 
seems too large to exclude an artifact in a resolution-convolution fitting. 
Thus the theoretical two-singlet scenario \cite{Bonville11} assuming 
the tetragonal distortion would be seriously modified 
to account for the present $E$-spectra and $T$-dependence, 
in which the crossover around 0.4~K could perhaps be ascribed 
to a structural transition. 

\begin{figure}
\begin{center}
 \includegraphics[width=0.45\textwidth]{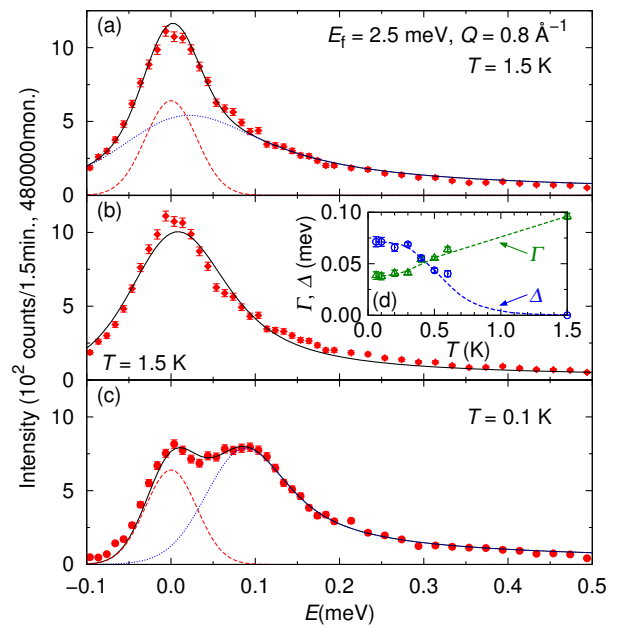}
\caption{
Results of fitting constant-$Q$ scan data at $T = 1.5$~K (a,b) 
and 0.1~K (c) shown in figure \ref{fig.2} to resolution 
convoluted $S(Q,E)$ of equation (\ref{eq.1}). 
We assume (a) $\Delta = 0$, and (b) $A = 0$ and $\Delta = 0$. 
Red dashed-lines and blue dotted-lines represent 
the resolution convolutions of the elastic and 
inelastic scattering of equation (\ref{eq.1}).
The black solid-lines are total fitted curves. 
(d) Temperature dependence of the fit parameters $\Gamma$ and $\Delta$. 
Lines are guides to the eye.
}
\label{fig.3}
\end{center}
\end{figure}
In order to parameterize the $E$ spectra shown in figure \ref{fig.2} 
as a function of temperature, we carried out fits of the data 
to a scattering function 
\begin{equation}
S(Q,E) = A\delta(E) + 
 \frac{B}{1 - e^{- E/k_{\text{B}} T}} 
 \sum_{\pm} \frac{\Gamma E}{(E \pm \Delta)^2  + \Gamma^2},
\label{eq.1}
\end{equation}
which is convoluted with a resolution function. 
The first and second terms of equation (\ref{eq.1}) represent 
elastic and inelastic scatterings, respectively. 
Typical results of the fitting are shown in figure \ref{fig.3}. 
For data at 1.5~K we tried to fit the spectrum using the 
Lorentzian function ($\Delta = 0$) for standard quasielastic 
scattering with ($A \neq 0$, figure \ref{fig.3}(a)) 
and without ($A = 0$, figure \ref{fig.3}(b)) the elastic component. 
These fits show that the elastic component, 20$\sim$30\% of 
the energy integrated ($E < 0.5$~meV) total intensity, cannot be 
neglected in this analysis. 
The elastic part consists of incoherent elastic scattering 
of Ti nuclei, multiple Bragg scattering, and 
magnetic scattering of Tb$^{3+}$ spins \cite{Gardner03}. 
In the present experiment, the separation of the elastic 
scattering from the inelastic scattering is difficult owing to 
the insufficient instrumental energy-resolution 
compared to the energy scale of 0.1 meV. 
We could not obtain a reasonable $T$-dependence of the 
parameter $A$ of equation (\ref{eq.1}), and 
assumed the same $A$ value for all temperatures. 
Resulting fits are shown in figures \ref{fig.3}(a) and (c).

The temperature dependence of the fit parameters 
$\Gamma$ and $\Delta$ are plotted in figure {\ref{fig.3}}(d). 
These parameters show the crossover behavior around 0.4~K. 
At 0.1~K the second term of equation (\ref{eq.1}) 
with $\Delta \sim 2 \Gamma$ represents a quasi-gapped 
continuum with a significant spectral weight at $E=0$. 
It should be noted that the very high resolution 
NSE data \cite{Gardner03} imply that $\sim 20$\% of 
the elastic component in equation (\ref{eq.1}) contains 
a magnetic contribution below 0.3~K.
Thus the magnetic spectrum, background subtracted 
equation (\ref{eq.1}), in the spin-liquid state probably has 
a two-peak structure at $E=0$ and 0.07 meV. 
This should be confirmed more directly using another 
spectrometer with a much higher resolution $\sim 10$~$\mu$eV 
or better in further work. 

\begin{figure}
\begin{center}
 \includegraphics[width=0.45\textwidth]{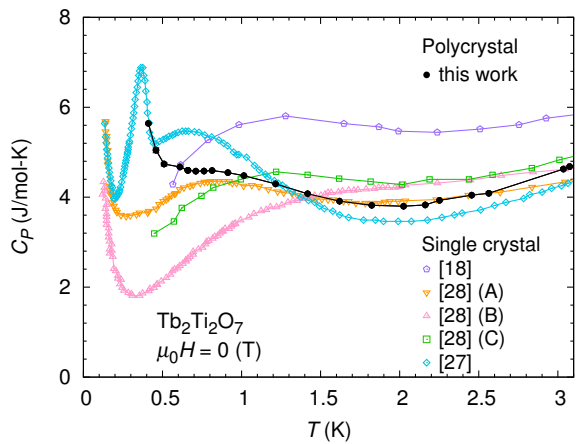}
\caption{
Temperature dependence of specific heat of polycrystalline and 
single-crystalline samples \cite{Gingras00,Chapuis09,Hamaguchi04}. 
}
\label{Cp}
\end{center}
\end{figure}
We measured the specific heat $C_{\text{P}}$ of the polycrystalline 
sample to check whether the crossover around 0.4~K can be 
observed in thermodynamic properties. 
The result is plotted in figure \ref{Cp} together with 
previous measurements. 
$C_{\text{P}}$ of the present work shows an upturn 
below 0.5~K, being consistent with the neutron data. 
We checked that the same upturn of $C_{\text{P}}$ 
is observed by a polycrystalline sample prepared in the same 
way as described in \cite{Gardner99,Gardner03}. 
One can also see differences of $C_{P}$ for 
crystalline samples \cite{Gingras00,Chapuis09,Chapuis10,Hamaguchi04}, 
demonstrating the difficulty of discussing 
experimental data taken on different single crystals, 
especially below 1 K. 
Control parameters of single crystals could be very 
small disorders, Tb/Ti site interchange, oxygen deficiency, 
or local stress built-in during single-crystal growth 
carried out using image furnaces. 
We speculate that the large differences of $C_{P}$ at low $T$ may 
imply that the system is located close to a quantum critical point 
which is affected by these hidden material parameters. 
We hope that the mechanism of the spin-liquid state can be 
explored further in studies on well-characterized single crystals. 

\noindent \textbf{4. Conclusion}

In summary, inelastic neutron scattering has been used to extend 
the work of \cite{Gardner99,Gardner03} and explore 
the spin-liquid state of polycrystalline Tb$_2$Ti$_2$O$_7$.
This system condenses into a quantum spin-liquid 
state from the paramagnetic state via a crossover around 0.4~K, 
where specific heat shows an anomaly. 
The energy spectrum in the spin-liquid state is a continuum 
with a double-peak structure at energies of 0 and 0.8~K. 
Its wave-vector dependence suggests that the spin-liquid 
state is brought about by many-body effects~\cite{Gingras00,Molavian07}.

\noindent \textbf{Acknowledgments}

We thank M. J. P. Gingras, R. Higashinaka, Y. Maeno, 
Y. Yasui, and S. Yonezawa for useful discussions. 
This work was supported by 
KAKENHI on Priority Areas ``Novel States of Matter Induced by Frustration'' 
and by the US-Japan Cooperative Program on Neutron Scattering. 

\bibliography{TTO}
\end{document}